\documentclass[10pt,letterpaper]{article}
\usepackage{opex3}
\usepackage{color}

\begin{document}

\title{A study of optical vortices inside the Talbot interferometer}

\author{Pituk Panthong$^1$, Sorakrai Srisuphaphon$^2$, Apichart Pattanaporkratana$^1$, Surasak Chiangga$^1$ and Sarayut Deachapunya$^{1,2,*}$}

\address{$^1$Department of Physics, Faculty of Science, Kasetsart University, Bangkok Province, 10900, Thailand\\
$^2$Department of Physics, Faculty of Science, Burapha University, ChonBuri Province, 20131, Thailand}

\email{$^*$sarayut@buu.ac.th} 



\begin{abstract}
The optical Talbot interferometer has been used to explore the topological charges of optical vortices. We recorded the self-imaging of a diffraction grating in the near-field regime with the optical vortex of several topological charges. Our twisted light was generated by a spatial light modulator (SLM). Previous studies showed that interferometric methods can determine the particular orbital angular momentum (OAM) states, but a large number of OAM eigenvalues are difficult to distinguish from the interference patterns. Here, we show that the Talbot patterns can distinguish the charges as well as the OAM of the vortices with high orders. Owing to high sensitivity and self-imaging of Talbot effect, several OAM eigenvalues can be distinguished by direct measurement. We assure the experimental results with our theory. The present results are useful for classical and quantum metrology as well as future implementations of quantum communications.
\end{abstract}



\section{Introduction}

Optical vortex or phase singularity is a twisted light which contains an information of phase and orbital angular momentum (OAM) of light~\cite{Allen1992}. It can be produced with several techniques~\cite{He1995,Kim1997,Masajada2001,Rotschild2004,Wei2009}. Among them, the use of a spatial light modulator (SLM) is reliable and simple~\cite{Ostrovsky2013}. The optical vortex has been experimentally used to perform both of classical and quantum optics. Far-field diffraction with vortex beams was exhibited for example the metrology~\cite{Senthilkumaran2012}, the evaluation of the topological charges of the vortex beams in the double slit experiment~\cite{Emile2014}, and the study of the interference pattern depending on the amount of OAM and the slit position in relation to the beam~\cite{Ferreira2011}. Trapping of particles can also be used from an optical vortex~\cite{Gahagan1996}. In quantum optics, the optical vortices of single photons were taken to study in Mach-Zehnder interferometer~\cite{Leach2002}, quantum entanglement~\cite{Mair2001}, and even quantum cryptography~\cite{Vaziri2002,Mirhosseini2015}. Characterization of OAM states of light has been a major technical challenge over the past decade. For the topological charge identification, the experiments were done not only the double slit diffraction but also the single slit diffraction~\cite{Ghaia2009}. However, this charge separation~\cite{Soskin1997} is hardly performed in the high vortex numbers~\cite{Ghaia2009}. Here, we show the use of optical near-field effect such as Talbot effect to separate the charge and the vortex numbers at higher orders since the sensitivity of the near-field Talbot effect is extremely high~\cite{Deachapunya2014}. The self-imaging of periodic structures in optical near-field diffraction can be observed when a parallel monochromatic light enters a diffraction grating and forms an interference pattern behind the grating. This effect is called the Talbot effect~\cite{Talbot1836}. The interference pattern produced from the Talbot effect has the period similar to the grating itself at the distance between the grating and the observation screen of one Talbot length, $L_{T}=d^{2}/\lambda$ as well as the rational multiples of Talbot length, $iL_{T}/j$, where $d$ is the grating period, $\lambda$ is the wavelength of the source, $i$ and $j$ are the integer number~\cite{Case2009,Srisuphaphon2015}. In our experiment, this distance was set to the multiples of this Talbot length and we will show that this scheme can clarify the charges and the OAM numbers of the vortices explicitly.

\section{Theory and method}

\begin{figure}[htb]
\centering\includegraphics[width=1\columnwidth]{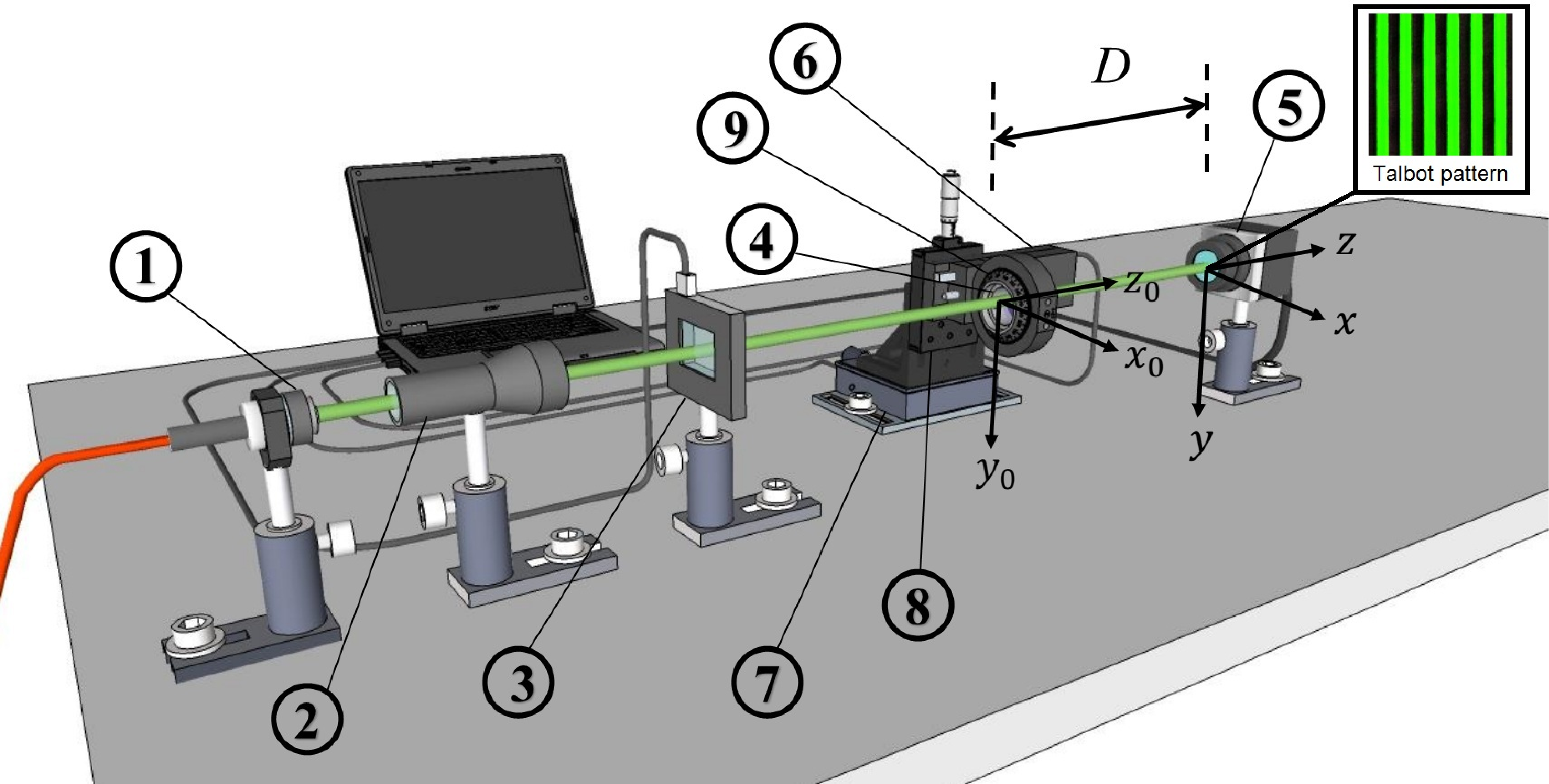}
\caption{(Color online) The Talbot setup with an optical vortex. (1): laser diode; (2): telescope; (3): SLM; (4): grating; (5): CCD camera; (6): first translation stage (z axis); (7): second translation stage (x axis); (8): third translation stage (y axis); (9): rotation grating holder. The inset shows a Talbot pattern at one Talbot distance recorded by the CCD camera with the OAM number or vortex number $l=0$. A spatial light modulator is used to produce optical vortices: the details described in the text.}
\label{Fig1}
\end{figure}

In this section, we present a short derivation of the Talbot effect involving an optical vortex. Assuming a Gaussian beam is propagating through a SLM along $z$ axis. The field distribution of an optical vortex order $l$ on $x_0y_0$ plane is given by
\begin{equation} \label{initial wave}
\psi(x_0,y_0,z_0=0)=(x_0+iy_0)^l \exp\{(-1/w^2)(x_0^2+y_0^2)\},
\end{equation}
where $w$ is the radius of the Gaussian beam. The vortex beam then incidents onto the grating at $z_0=0$ (Figure~\ref{Fig1}). The grating transforms the wave to be
\begin{equation} \label{initial G wave}
\psi_1(x_0,y_0,+0)=\sum_{n}A_n exp\{i n k_d x_0\}\psi(x_0,y_0,-0),
\end{equation}
where $n=0,\pm1,\pm2,...$, and $k_d=2 \pi /d $ with $d$ is the grating period. The factor $A_n=sin(n \pi f) / n \pi$  is the components of the Fourier decomposition of the periodic for the grating  with an opening fraction $f$ ~\cite{Case2009}.

In order to consider the effect of an optical vortex to the Talbot vortex, we apply the Huygens-Fresnel integral to find the field distribution on the screen at distance $D$ behind the grating. Let $xy$ plane is the transverse plane at the screen (Figure~\ref{Fig1}), the field distribution $\psi_2(x,y)$ is given by

\begin{eqnarray} \label{wave1}
\psi_2(x,y)=\frac{ik}{2 \pi D} \int_{-\infty}^{\infty}\int_{-\infty}^{\infty}dx_0dy_0   \sum_{n}A_n exp\{in k_d x_0 - ik(D+\frac{(x-x_0)^2}{2D}+\frac{(y-y_0)^2}{2D})\}
\psi_1(x_0,y_0,+0),
\end{eqnarray}
where $k=2\pi / \lambda$ is the magnitude of the wave vector with wavelength $ \lambda$. According to the field $\psi_1$ in Eq.(\ref{initial G wave}), the field $\psi_2(x,y)$ can be written as

\begin{eqnarray} \label{wave2}
\psi_2(x,y)=\frac{ik}{2 \pi D}\exp\{-i k D-\frac{ik}{2D}(x_0^2+y_0^2)\}  \sum_{n}A_n P_n(x,y),
\end{eqnarray}
where
\begin{eqnarray} \label{I1}
P_n(x,y)=\int_{-\infty}^{\infty}\int_{-\infty}^{\infty}dx_0dy_0(x_0+iy_0)^l \exp\{ -C(x_0^2+y_0^2)+ i(k/D)(xx_0+yy_0)+ i n k_d x_0 \},
\end{eqnarray}
and $C=(\frac{1}{w^2}+\frac{ik}{2D})$. The above integral can be evaluated by applying the procedure as shown in Ref.\cite{Kumar2011}. Namely, letting $X=(k/D) x$, $Y=(k/D) y$ and using the identity $(-i)^l(\frac{\partial}{\partial X}+i\frac{\partial}{\partial Y})^l\exp\{i(x_0X+y_0Y)\}=(x_0+iy_0)^l
\exp\{i(x_0X+y_0Y)\}$, the integral can be done analytically by using $\int_{-\infty}^{\infty}du \exp (-au^2-2bu)=\sqrt{\pi/a}\exp (b^2/a)$, and the result as,
\begin{eqnarray} \label{I2}
P_n(x,y)=(-i)^l(\frac{\partial}{\partial X'}+i\frac{\partial}{\partial Y'})^l (\frac{\pi}{C})\exp\{-\frac{(X'^2+Y'^2)}{4C}\},
\end{eqnarray}
here we let $X'=X+n k_d$ and $Y'=Y$. For evaluating the partial derivative, we use $(\frac{\partial}{\partial X'}+i\frac{\partial}{\partial Y'})^l \equiv (2 \frac{\partial}{\partial \rho^\ast})^l
$ with $\rho=X'+iY'$, Eq.(\ref {I2}) is reduced to
\begin{eqnarray} \label{I3}
P_n(x,y)=(\frac{\pi}{C})\big(\frac{i \rho}{2C}\big)^l \exp\{-\frac{\rho\rho^\ast}{4C}\}.
\end{eqnarray}
Substituting Eq.(\ref{I3}) into Eq.(\ref{wave2}) with the defined variables, we obtain the field distribution on the screen as

\begin{eqnarray} \label{wave3}
\psi_2(x,y)&=&\big(\frac{i k}{2CD}\big)^{l+1}\exp\{-i k [D+\frac{(x^2+y^2)}{2D}]\}
\nonumber
\\ && \sum_{n}A_n \big( x+\frac{nk_dD}{k} +i y \big)^l\exp\{\big(-\frac{k^2}{4CD^2} \big)\big[ (x+\frac{nk_dD}{k})^2+y^2 \big]\}.
\end{eqnarray}
In analogy, the optical vortex with minus $l$ can be obtained by replacing the initial wave (Eq.(\ref{initial wave})) with $\psi(x_0,y_0,z_0=0)=(x_0-iy_0)^{|l|} \exp\{(-1/w^2)(x_0^2+y_0^2)\}$. So, the intensity distribution $\psi_2^\ast\psi_2$ may be written in the form

\begin{eqnarray} \label{intensity}
I_{\pm l}(x,y)&=&\big(\frac{k^2}{4D^2C^\ast C}\big)^{|l|+1}\sum_{n,n'}A_n A_{n'}\big( x+\frac{nk_dD}{k} \pm iy \big)^{|l|}\big( x+\frac{n'k_dD}{k} \mp iy \big)^{|l|}
\nonumber
\\ && \exp\{\big(-\frac{k^2}{4CD^2} \big)\big[ (x+\frac{nk_dD}{k})^2+y^2 \big]\}
\nonumber
\\ && \exp\{\big(-\frac{k^2}{4C^\ast D^2} \big)\big[ (x+\frac{n'k_dD}{k})^2+y^2 \big]\},
\end{eqnarray}
where the subscribe $\pm l$ denote the optical vortex of order $l>0$ and $l<0$, respectively.

\section{Experiment}

A 5 mW green diode laser ($\lambda=532$nm) is used in our experiment as a coherent source for the Talbot interferometer. The laser source is expanded by an optical telescope to a diameter of about 20 mm in order to cover the whole grating which makes a clean shape of the Talbot image. Our optical vortex is generated by using the SLM (LC2012, Holoeye Photonics AG with resolution of 1024x768 pixels) (Figure~\ref{Fig1}). A series of gray-level images is loaded and displayed onto the SLM screen for producing the vortices (Figure~\ref{Fig3}-~\ref{Fig5} (a), and (b)). The azimuthal phase varies from $0$ to $2\pi$ with 8-Bit (256 grey levels) level at 532 nm of the laser source. Afterward, the vortex beam is illuminated a diffraction grating (chromium on glass, Edmund Optics inc., $d=200\mu m,f=0.5$). The distance between the SLM and the grating is about 4.5 meters. An interference pattern or Talbot pattern is detected with a CCD camera (DCU223C, Thorlabs) acting as a screen at $D=5L_{T}$ by moving the grating with the first translation stage (MTS50/M-Z8E, resolution 1.6 $\mu$m, Thorlabs). The grating is also movable in the x axis and y axis by using the second translation stage (Z812B, Thorlabs) and the third translation stage (PT1/M, Thorlabs), respectively. The grating can be moved in these transverse directions for adjusting the vortex beam to the center of the grating. The rotation grating holder (LCRM2/M, Thorlabs) is helped to align the interference pattern to the CCD camera.

\begin{figure}[htb]
\centering\includegraphics[width=0.5\columnwidth]{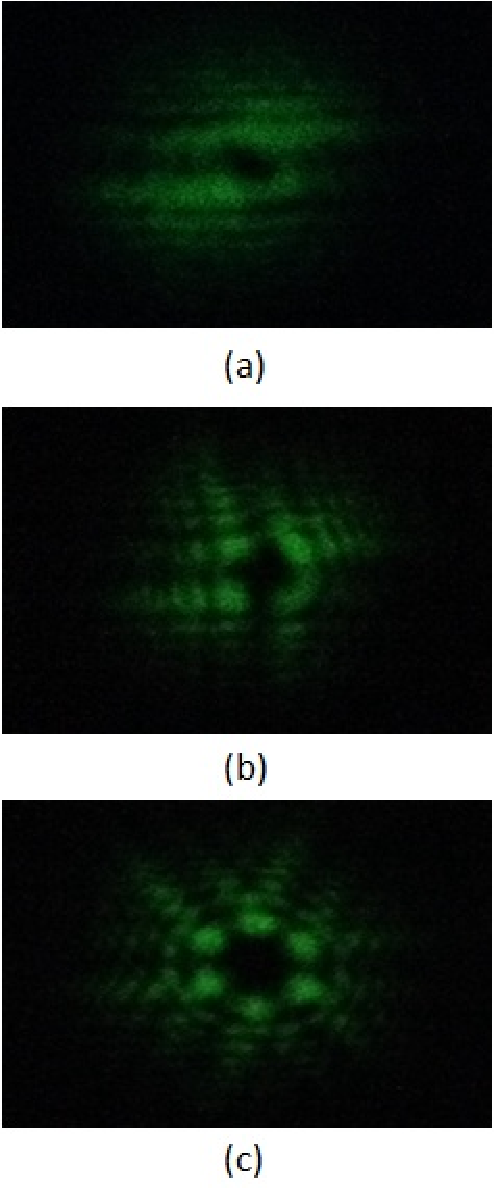}
\caption{(Color online) The optical vortices produced from our SLM with the topological charges of $l=1$ (a), $2$ (b), $3$ (c).}
\label{Fig2}
\end{figure}

\section{Results and discussions}

\begin{figure}[htb]
\centering\includegraphics[width=1\columnwidth]{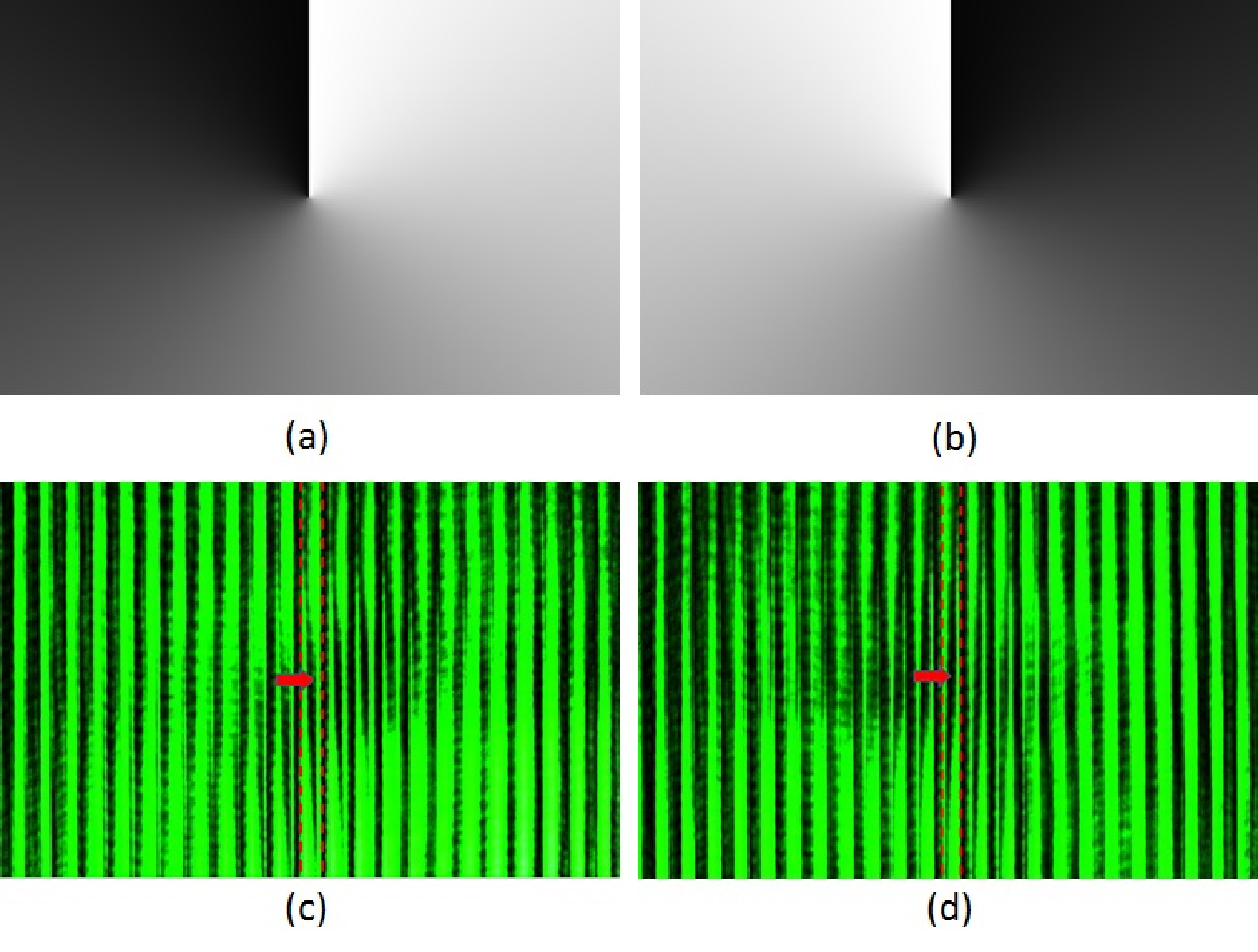}
\caption{(Color online) The gray-level patterns used for producing the optical vortices of charge 1 (a), and -1 (b) as well as the Talbot patterns modulated with the vortices of charge 1 (c), and -1 (d). The dashed lines represent the boundary of the vortex within one bright stripe and the arrows point the dark stripes appeared in the center of the bright fringes.}
\label{Fig3}
\end{figure}

The main goal of this research is to use the Talbot interferometer as a tool to explore the OAM of light and also to distinguish the topological charges of the optical vortices. Since the difference charge gives also the magnitude and the opposite direction of the twisted light, the trail of vortex inside the Talbot pattern should be revealed obviously. We first produced (Figure~\ref{Fig2}) and performed the vortices with orbital $l=1$ and $l=-1$ in the Talbot setup. The phase modulations of both the vortex itself and the near-field diffraction present the modification of fringe patterns as shown in Figure~\ref{Fig3}. The interference patterns are splitting. The direction of twisted light is exhibited. For orbital $l=1$, the upper parts of the pattern are shifted to the right and the lower parts are shifted oppositely as expected and one could observe the dark stripes appeared in the center of the bright fringes as indicated with the arrow in Figure~\ref{Fig3}(c). For the orbital $l=-1$, the situation is reversed (Figure~\ref{Fig3}(d)).

\begin{figure}[htb]
\centering\includegraphics[width=1\columnwidth]{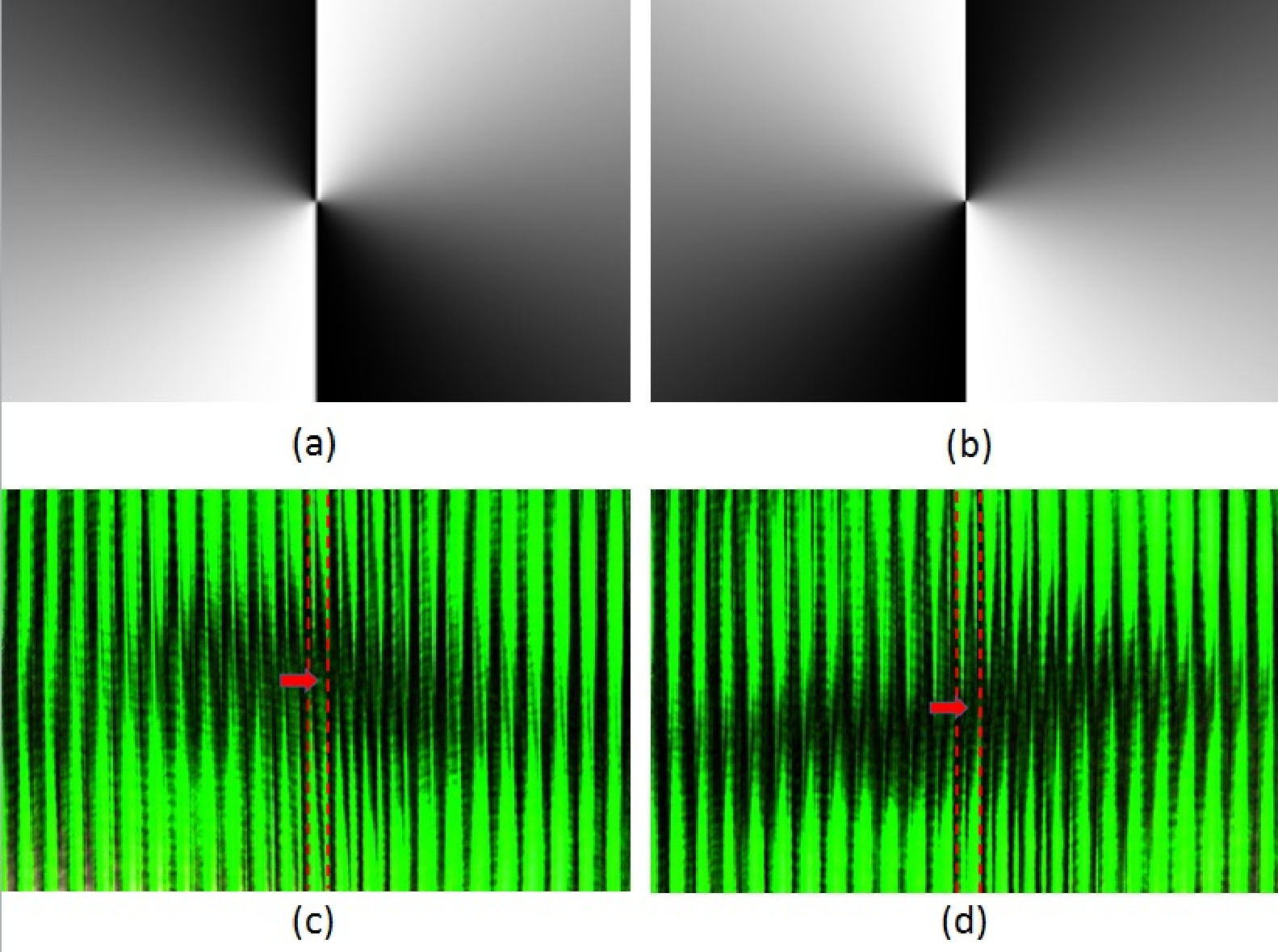}
\caption{(Color online) The gray-level patterns used for producing the optical vortices of charge 2 (a), and -2 (b) as well as the Talbot patterns modulated with the vortices of charge 2 (c), and -2 (d).The dashed lines represent the boundary of the vortex within one bright stripe and the arrows point the dark stripes appeared in the center of the bright fringes.}
\label{Fig4}
\end{figure}

\begin{figure}[htb]
\centering\includegraphics[width=1\columnwidth]{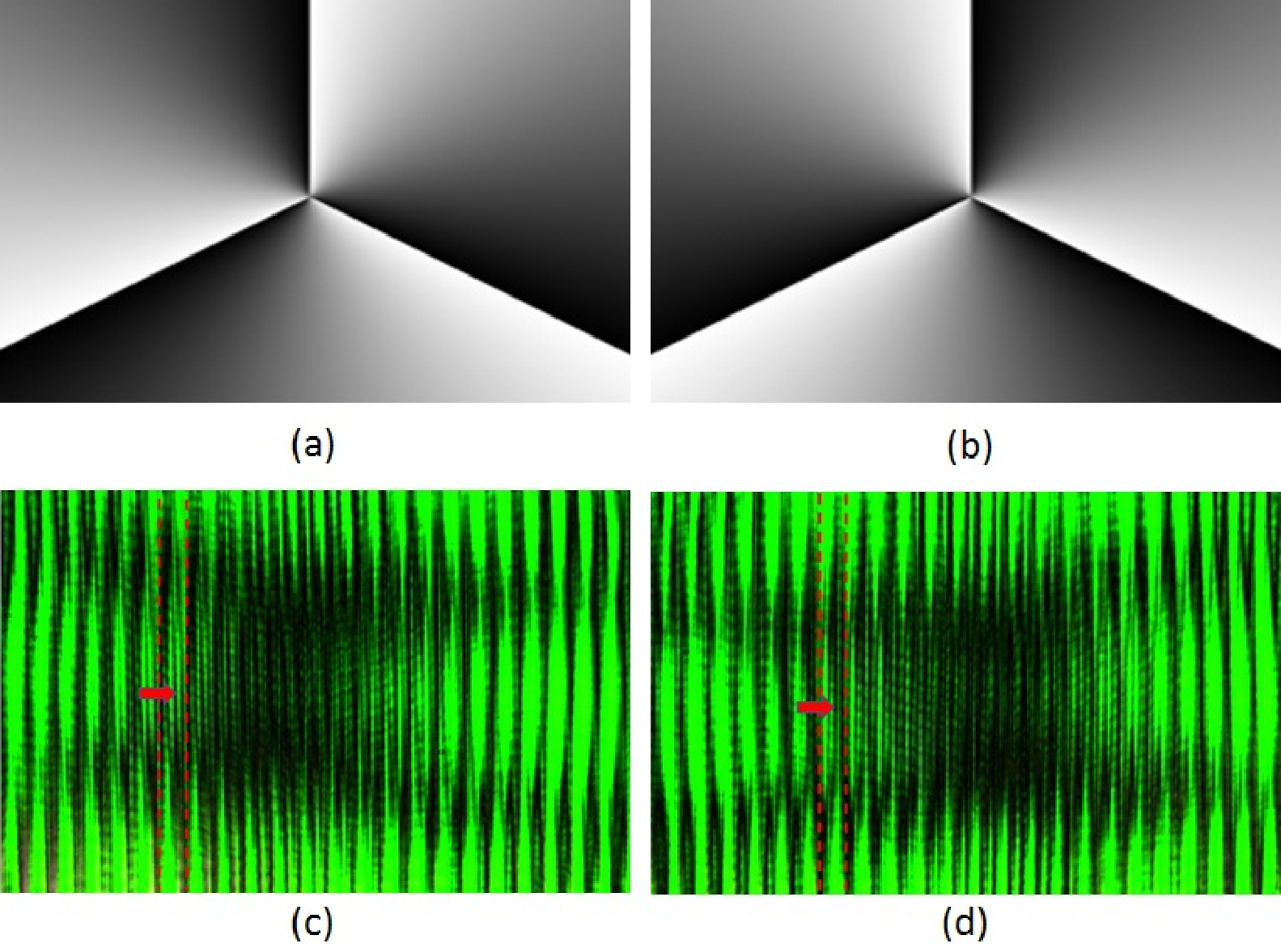}
\caption{(Color online) The gray-level patterns used for producing the optical vortices of charge 3 (a), and -3 (b) as well as the Talbot patterns modulated with the vortices of charge 3 (c), and -3 (d).The dashed lines represent the boundary of the vortex within one bright stripe and the arrows point the dark stripes appeared in the center of the bright fringes.}
\label{Fig5}
\end{figure}

\begin{figure}[htb]
\centering\includegraphics[width=1\columnwidth]{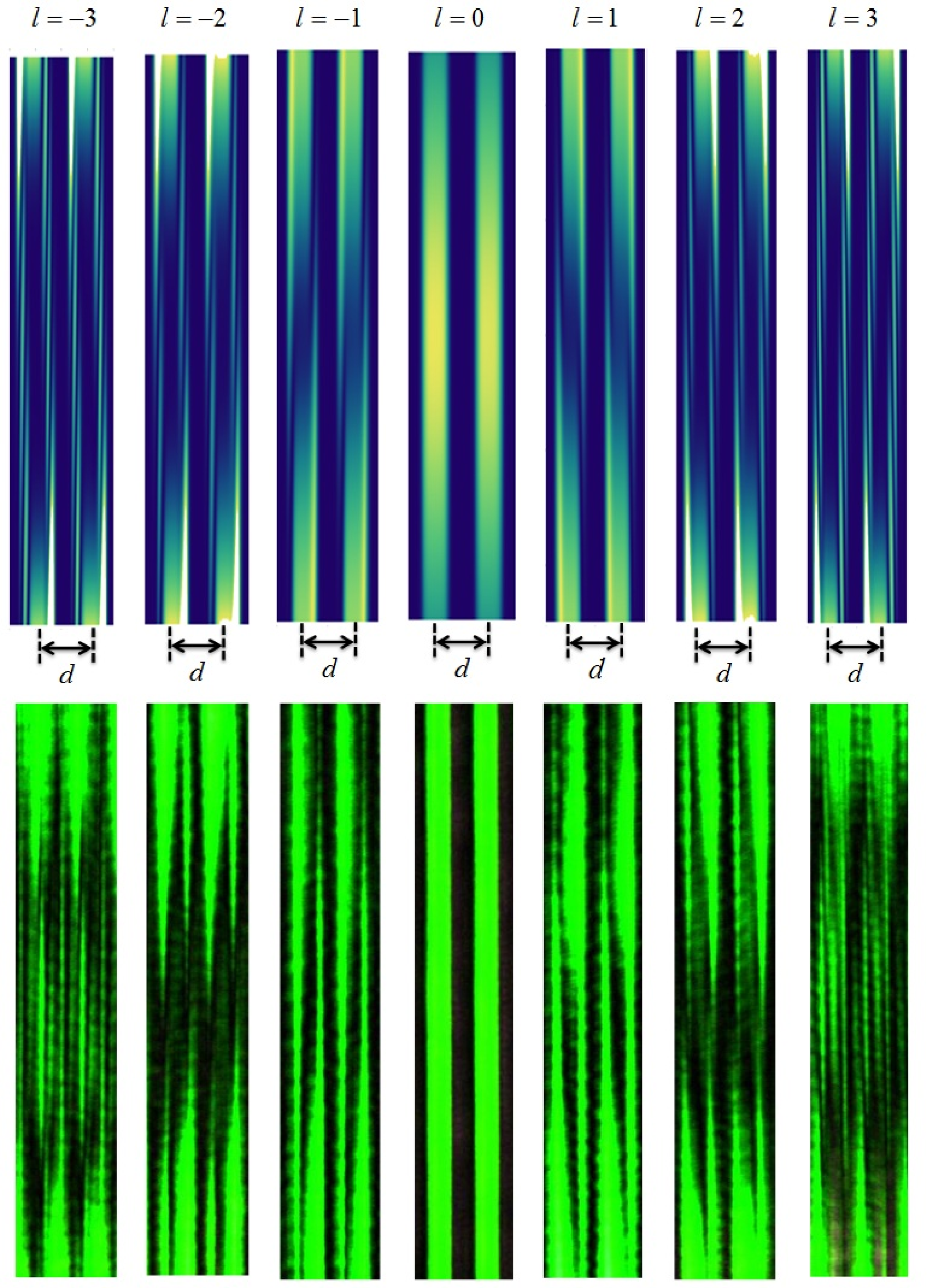}
\caption{(Color online) The comparison between the simulations according to Eq.(\ref{intensity}) and the experimental results as shown in Figure~\ref{Fig3}, ~\ref{Fig4}, and ~\ref{Fig5} for $l=-3, -2, -1, 0, 1, 2$, and $3$. They are in good agreement with each other. $l$ labels the vortex numbers, and $d$ means the grating or fringe period.}
\label{Fig6}
\end{figure}

Subsequently, we continued the experiment with four more charges of $l=2$, $-2$, $3$, and $-3$. Figure~\ref{Fig4} and ~\ref{Fig5} clarify the explanation above apparently. For higher orbital numbers, the more fringe tilts occur and also the major dark spots could be seen larger. We assure the results with our theoretical approach. Figure~\ref{Fig6} shows the comparison between the simulations according to Eq.(\ref{intensity}) and the experimental results as shown in Figure~\ref{Fig3}, ~\ref{Fig4}, and ~\ref{Fig5} for $l=-3, -2, -1, 0, 1, 2$, and $3$. $l=0$ means the Talbot pattern with Gaussian beam of $w=4$mm. The calculations were done with the truncated sum at $n=\pm 25$. The experimental results are in good agreement with the simulation nicely. According to the results, the dark stripes inside each of bright fringes can be used to determine the OAM number ($l$). For example, for $l=1$ and $l=-1$, there is only one dark (tilt) stripe inside each bright fringe pattern and for $l=3$ and $l=-3$, there are three dark stripes for instance (Figure~\ref{Fig6}). This appears at both simulations and experiments. Here, we show that the Talbot pattern can be used to determine magnitude and sign of the topological charge for high orbital numbers as soon as the size of the vortex is smaller than the detection area.

\begin{figure}[htb]
\centering\includegraphics[width=1\columnwidth]{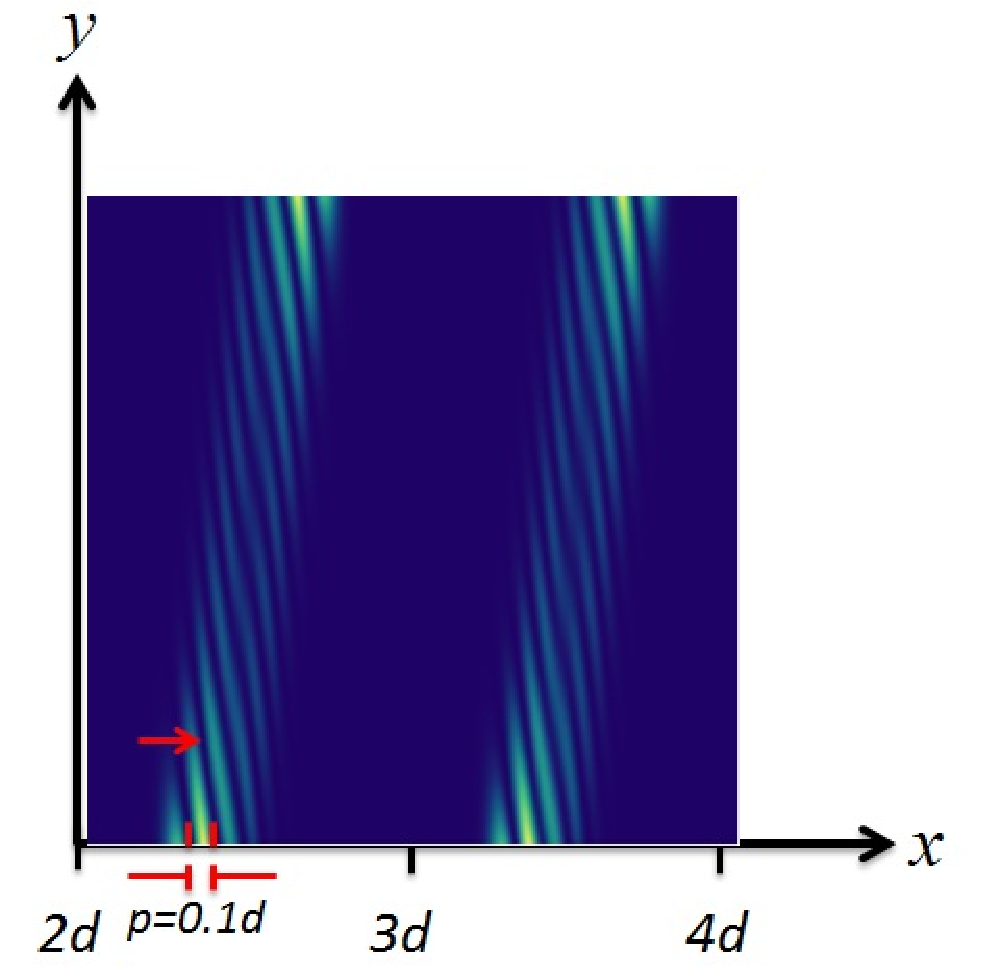}
\caption{(Color online) The calculation of Talbot optical vortex for the orbital number of $l=10$. The simulation were done with the grating period ($d$) of 200 $\mu$m, the laser wavelength of 532 nm, and the opening fraction ($f$) of 0.1. It shows ten dark stripes (one indicated by the arrow) inside each of bright fringes as described in the text. $p$ is the width of bright fringes without the vortex (see text).}
\label{Fig7}
\end{figure}

Although our scheme is adequate for characterization of the OAM states of light, we nevertheless discuss the possibility to improve the resolution of our method. The calculations with different conditions were performed with several grating periods ($d$) and opening fractions ($f$). We conclude that for such high OAM numbers, the smaller opening fraction provides a better resolution while the grating period is not crucial. Figure~\ref{Fig7} presents an example of the simulation of the Talbot pattern with optical vortex of $l=10$. This simulation was done with $d=200\mu$m, $f=0.1$, and $\lambda=532$ nm. The result also shows ten dark stripes inside each of bright fringes as described above. It also appears that each bright fringe has the width larger than a fringe without vortex ($p$) (see Figure~\ref{Fig7}) because the vortex splits the upper half of the bright fringes to the right and the lower half to the opposite direction. The smaller opening fraction ($f=0.1$) has more advantages than the larger one ($f=0.5$) because each of bright fringes and its nearest neighborhood can not disturb each other since the space between them is larger.

According to the present results, we therefore summarize that the near-field Talbot effect can provide a good tool to realize an optical vortex and has a possibility to use as a method for quantum information technology since we can distinguish two orthogonal vortex states with $+l$ and $-l$.

\section{Conclusion}

The Talbot experiment was performed with the vortex of light as a source. The experiments were done with several vortex numbers and topological charges. We used the near-field Talbot effect as a tool to distinguish the different orbital numbers and also the opposite charges. The Talbot effect with optical vortex may be applied to the field of quantum information if we perform with a single photon source since we can distinguish the opposite charges of an optical vortex.

\section{Acknowledgments}

 S.D. acknowledges the support grant from the office of the higher education commission, the Thailand research fund (TRF), and Faculty of Science, Burapha university under contract number MRG5380264. We would like to thank Dr. Boonlit Krunavakarn and Dr. Nupan Kheaomaingam for the useful discussions.

\end{document}